# Snapshot multispectral imaging using a filter array

Kazuma Shinoda[1]

1 Graduate School of Engineering, Utsunomiya University
7-1-2 Yoto, Utsunomiya, 321-8585 Japan
E-mail: shinoda@is.utsunomiya-u.ac.jp

**Abstract** A multispectral filter array (MSFA) is one solution for capturing a multispectral image (MSI) in a single shot at low cost. We introduce our optimization method of the spectral sensitivity of the MSFAs and demosaicking, and show a new prototype filter array for snapshot imaging based on a photonic crystal.

## 1. Introduction

Multispectral images (MSIs) have been studied for many fields, such as remote sensing, medical applications, and color reproduction [1-3]. Some multispectral cameras have also been proposed [4, 5], but practical use of multispectral imaging has generally been limited because of the extremely high cost of applications and their lack of portability. Multispectral filter arrays (MSFAs) are a possible solution to this problem. The architecture of an MSFA has a different spectral filter for each pixel of an image. Images captured by an MSFA have only one value for each pixel, but a full-resolution MSI can be obtained by recovering the missing spectrum information. This recovering process is referred to as *demosaicking*. Various filter array patterns and demosaicking methods [6-21] have been proposed for improving demosaicked image quality. Because the number of non-measured pixels in an MSFA is larger than that in an RGB Bayer color filter array, a more appropriate MSFA pattern and an accurate demosaicking method are required.

In this paper, we present the state-of-the-art of MSFA, and introduce our designing and demosaicking approaches. Then, we show a photonic-crystal-based filter array as an example of a fabricated MSFA.

## 2. Multispectral filter array design and demosaicking

A few examples of MSFAs are shown in Fig. 1. Brauers et al. [6] proposed a six-band MSFA arranged in 3 x 2 pixels in a straightforward manner for fast linear interpolation. Their demosaicking method first applies bilinear interpolation to each band. Inter-band correlations are then corrected by smoothing inter-band differences. Yasuma et al. [7] designed a seven-band MSFA composed of three primary color filters and four secondary color filters. A simple four-band MSFA was proposed by Aggarwal et al. [8], who presented a versatile demosaicking method based on an $l_1$-norm minimization problem. Monno et al. [12] proposed a five-band MSFA and determined that the sampling density of G-band data was higher than that of the other spectral bands because the human eye is more sensitive to the G-band compared to other spectral bands. We proposed an optimization method for MSFA arrangement based on a new metric called the interpolation quality metric (IQM) [13], which is shown to be proportional to the PSNR of the demosaicked image. We also improved and generalized this metric to average nearest-neighbor distance (ANND) [14], and showed the optimized pattern for recovering spectra of visible wavelengths. These methods do not require any training data.

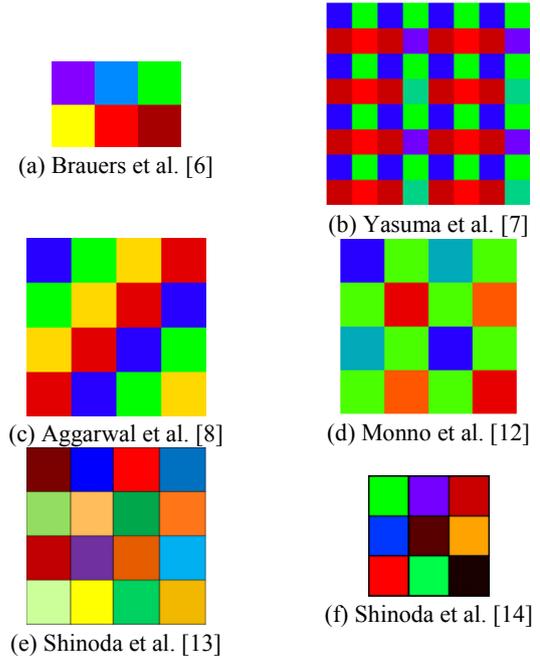

(a) Brauers et al. [6]
(b) Yasuma et al. [7]
(c) Aggarwal et al. [8]
(d) Monno et al. [12]
(e) Shinoda et al. [13]
(f) Shinoda et al. [14]

Fig. 1. Examples of MSFAs.

For demosaicking, Wang et al. [15] proposed a wavelet-based demosaicking method. This method interpolates a higher spatial subband coefficient from other spectral bands. Miao et al. [16] proposed a binary-tree-based demosaicking method. They assume that the number of sampled pixels differs at each band, and the most dense spectral band is interpolated with high priority than the other bands. Mihoubi et al. [17] proposed a demosaicking method based on the correlation between each spectral band and a pseudo-panchromatic image (PPI), which represents the average image over all bands. We proposed a demosaicking method using vectorial total variation (VTV) regularization [18]. This process is regarded as inverse problem of the image observation model, and the reconstructed image is estimated by minimizing the VTV as a regularization term under the constraint.

Many conventional studies assumed that the spectral sensitivity of each filter was a narrowband filter (sometimes delta function). However, some studies have attempted to change this concept. Hirakawa et al. [19] suggested that panchromatic filters can prevent aliasing in mosaicked image, which implies that multiplexing broadband spectral filters can facilitate anti-aliasing in MSFAs. Jia et al. [20] designed the Fourier spectral filter array (FSFA), which comprises sinusoidal broadband spectral filters. The FSFA can reduce aliasing based on Fourier transform spectroscopy. They also proposed a linear modulation demosaicking method for the FSFA.

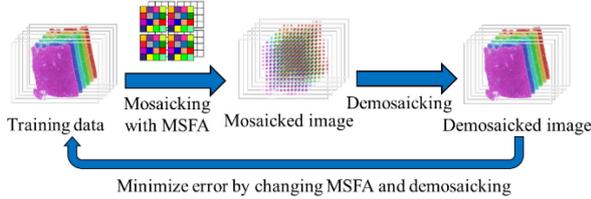

Fig. 2. Designing MSFA and demosaicking using full-resolution MSI [21].

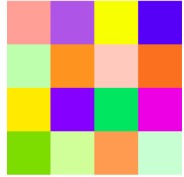

Fig. 3. Optimized MSFA using pathological images [21].

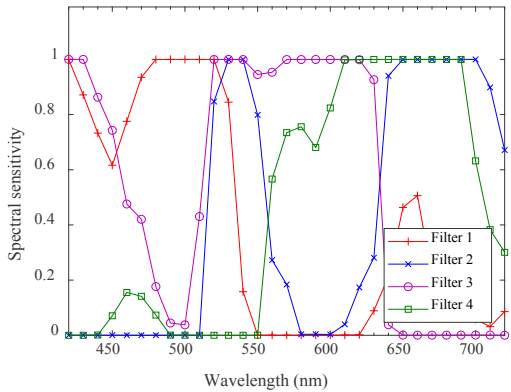

Fig. 4. Optimized spectral sensitivity (Filter 1 to 4) [21].

Beside these approaches, we proposed a joint optimizing method for the spectral sensitivity of an MSFA and the demosaicking with using a training data, and pathology-specific multispectral imaging is proposed [21]. When full-resolution MSIs, which are not mosaicked, are employed, the MSFA pattern and the demosaicking matrix can be optimized by minimizing the error between the original and demosaicked images during a simulation, as shown in Fig. 2. The proposed method first assumes that the mosaicking process using an MSFA corresponds to a linear system, and that the demosaicking process is regarded as an inverse problem of the linear model. The original image is mosaicked using a random MSFA through a simulation; subsequently, the mosaicked image is recovered using a Wiener estimation matrix. The spectral sensitivities of the MSFA are optimized iteratively using an interior-point approach to minimize the errors between the original and demosaicked MSIs. The Wiener estimation matrix is also optimized according to the changes in the spectral sensitivity.

The optimized filter pattern is shown in Fig. 3, and the part of the spectral sensitivities are shown in Fig. 4. Note that each pixel in Fig. 3 is colored according to the sRGB color reproduction under an illuminant of D65. Each filter of the optimized MSFA can be seen as a mixture of colors because the spectral sensitivity functions become relatively broad, and the appearance of these filters is slightly different compared with that of the conventional MSFAs in Fig. 1. The proposed method achieved approximately a 4 to 6 dB improvement in peak signal-to-noise ratio compared to the Monno MSFA [12]. The manufacturing costs of such a highly complicated MSFA will be examined in future.

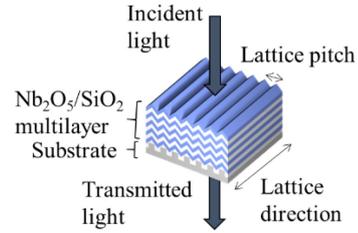

Fig. 5. Conceptual chart of photonic crystal.

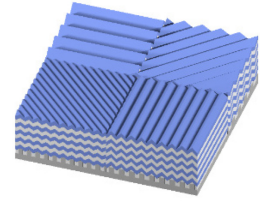

Fig. 6. Conceptual chart of photonic crystal filter array for snapshot multispectral polarization imaging.

### 3. MSFA fabrication using a photonic crystal

In this section, we introduce a photonic-crystal-based filter array as a fabrication example of MSFA. Many conventional studies assume to use a narrow bandpass filter for MSFA, but we utilizes various sensitivities produced by a photonic crystal.

A photonic crystal is an optical nanostructure with a period comparable to the wavelength of light. The transmitted spectrum is changed depending on the period. In particular, Ohtera et al. [22, 23] proposed a photonic crystal filter (PhCF), which is a thin-film wavy multilayer structure that is fabricated by the autocloning method based on a radio frequency (RF) bias sputtering process. Autocloning is a method to fabricate a multilayered-type photonic crystal based on lithography and sputtering [24, 25]. First, a lattice pattern is prepared on a substrate by lithography and dry etching. Next, by stacking multiple dielectric films on the substrate using the RF bias sputtering process, a structure that has refractive index modulation in both horizontal and vertical directions can be obtained. According to this method, by changing the period of lattice (in-plane lattice constant) on the initial substrate from one position to another, it is possible to fabricate multiple photonic crystal regions with different horizontal lattice structures and common vertical index profiles using a single sputtering process. An example of the fabricated two-dimensional multilayer wavy structure is shown in Fig. 5. As the transparent spectral sensitivity can be controlled by changing the lattice pitch of the wavy structure, an MSFA can be obtained by changing the lattice pitch at each pixel separately.

Although the photonic crystal with 2D wavy structure was an appropriate device for MSFAs, we further improved this structure for snapshot multispectral "and polarization" imaging [26]. The PhCF has a different spectral sensitivity in transverse electric (TE) and transverse magnetic (TM) polarization modes due to form birefringence. Y. Ohtera et al. [24, 25] used only one of these modes to fabricate edge filters by attaching an additional linear polarization filter. In contrast, we focused on the transmittance difference in TE and TM modes and utilized both modes for multispectral polarization imaging. A multispectral polarization filter array can be obtained by changing the lattice pitch and direction at each pixel separately as shown in Fig. 6.

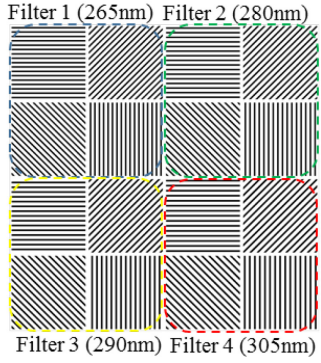

Fig. 7. Filter array pattern and lattice pitch (4 x 4 pixel).

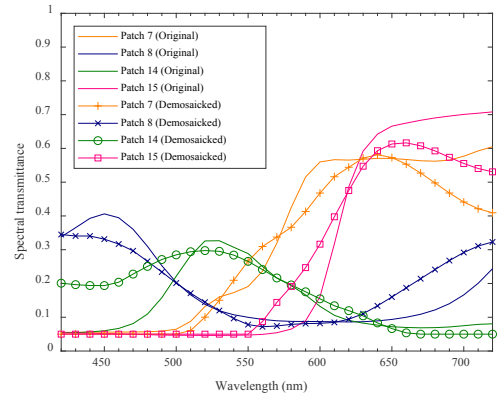

Fig. 9. Spectral reflectance comparison of color chart.

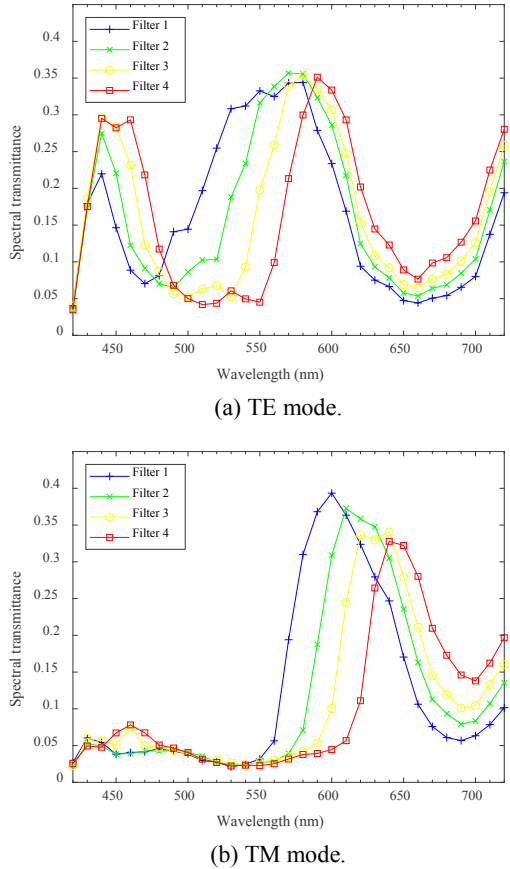

(a) TE mode.

(b) TM mode.

Fig. 8. Spectral transmittance of filter array.

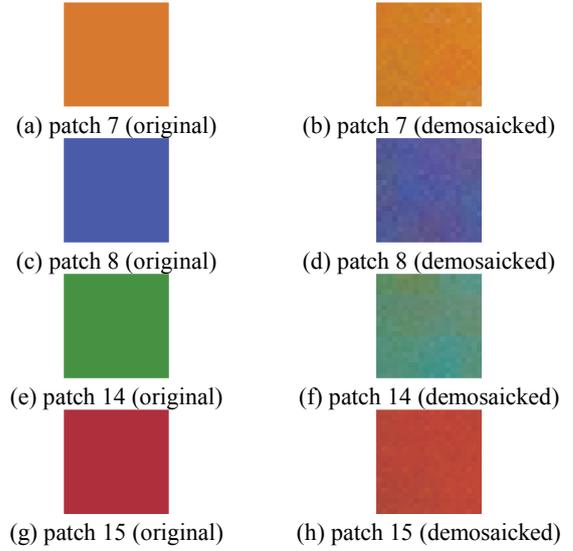

(a) patch 7 (original)   (b) patch 7 (demosaicked)

(c) patch 8 (original)   (d) patch 8 (demosaicked)

(e) patch 14 (original)  (f) patch 14 (demosaicked)

(g) patch 15 (original)  (h) patch 15 (demosaicked)

Fig. 10. RGB image of color chart.

We designed a multispectral polarization filter array as shown in Fig. 7. Fig. 7 is a lattice pattern on a substrate in the lithography process. The proposed filter array has 265, 280, 290, and 305nm pitches as filters 1, 2, 3, and 4, with each of them having four lattice directions, 0, 45, 90, and 135 degrees. The total of filter types is 4 x 4 = 16 pixels. The filter array is arranged periodically in horizontal and vertical directions until 100 x 100 pixels. The lithography process, sputtering process, material, thickness, and other parameters of multilayer structures can be referred to in [24].

Fig. 8 shows the measured spectral transmittance of filters. The fabricated filter has different spectral transmittances in TE and TM modes because of optical anisotropy. Moreover, because of changing lattice pitches, the spectrum peak moves from shorter to longer wavelength as shown from filters 1 to 4. From the results, it is clear that the fabricated multispectral polarization filter array has a different spectral and polarization transmittance property at each pixel.

However, it requires the demosaicking process to recover a multispectral polarization image from the captured grayscale image because various spectral and polarization components are mixed in one pixel. The capturing process is considered as a linear model and various images are recovered by solving the inverse problem of the linear model as well as [21].

We attached the fabricated filter array to a monochrome imager (ICX205AL, SONY) by ultraviolet curing. For alignment, we used a manipulator while viewing a captured monochrome image in real time. In the UV curing process, the filter array was moved on the surface of the imager using a manipulator. We exposed UV light at a position which maximizes the contrast of the captured mosaic pattern. We used an 8-bit USB camera casing (ARTCAM-150P5-BW-WOM, ARTRAY) for snapshot imaging. The exposure time of the camera was set to 80 ms. The recovering target wavelength was in the range of 420-720 nm at 10 nm interval (L = 31 bands); the D65 standard illuminant was used for RGB reproduction. In the following experiments, the spectrum reflectance was compared with an original spectrum in color chart (ColorChecker Classic, x-rite).

The results of non-polarized spectral reflectance demosaicked from a captured grayscale image are shown in Fig. 9. Here, we compare the original measured spectrum using a spectrometer (USB2000+, OceanOptics) and the demosaicked spectrum where the patch index is 7, 8, 14, and 15. From the comparison, the form of the demosaicked spectrum approximately matches the original spectrum. However, major difference can be seen in wavelengths shorter than 500 nm of patch 14 and wavelengths longer than 650 nm of all patches. For visual comparison, we show the reproduced RGB image from the original spectrum and demosaicked non-polarized RGB image in Fig. 10. The visual colors of demosaicked RGB are close to the original spectrum; in particular, there is almost no difference in patches 7 and 8. The color of patches 14 and 15 is a little low in the demosaicked results.

It is possible to reduce the demosaicked error by increasing the number of filter types. We designed only four types of transmittance (i.e., four types of lattice pitch) in Fig. 7; however, each filter has a similar transmittance curve in few wavelengths (particularly over 650 nm). This similarity may weaken the linear independence with each filter and may lead to degradation of demosaicking quality. It is easy to increase the number of filters in our method; therefore, a larger number of different lattice pitches should be used to improve the demosaicking quality in the future.

## 4. Conclusion

This paper focused on an MSFA architecture for snapshot multispectral imaging. We presented some conventional MSFA designing and demosaicking methods, and introduced our approaches. In particular, we showed a photonic-crystal-based filter array as an example of fabricated MSFAs. MSFA is suitable for hand-held and real-time multispectral imaging applications, therefore, it can be expected that snapshot technologies with MSFAs will become more important in many fields using MSIs.